\documentclass[twocolumn]{aastex63}

\newcommand{\ci}{C\,{\sc i}}
\usepackage{epsfig}
\usepackage{amsmath}

\received{2019 December 18}
\revised{2020 January 2}
\accepted{2020 January 2}
\submitjournal{ApJL}

\shorttitle{[\ci] to molecular gas mass conversion at high redshift}
\shortauthors{Heintz \& Watson}

\begin{document}

\title{\sc Direct measurement of the [\ci] luminosity to molecular gas mass conversion factor in high-redshift star-forming galaxies}

\correspondingauthor{Kasper E. Heintz}
\email{keh14@hi.is}

\author[0000-0002-9389-7413]{Kasper E. Heintz}
\affiliation{Centre for Astrophysics and Cosmology, Science Institute, University of Iceland, Dunhagi 5, 107 Reykjav\'ik, Iceland}

\author[0000-0002-4465-8264]{Darach Watson}
\affiliation{Cosmic Dawn Center (DAWN), Denmark}
\affiliation{Niels Bohr Institute, University of Copenhagen, DK-2100 Copenhagen, Denmark}

\begin{abstract}

The amount of cold, molecular gas in high-redshift galaxies is typically inferred from proxies of molecular hydrogen (H$_2$), such as carbon monoxide (CO) or neutral atomic carbon ([\ci]) and molecular gas mass conversion factors. The use of these proxies, however, relies on modeling and observations that have not been directly measured outside the local universe. Here, we use recent samples of high-redshift gamma-ray burst (GRB) and quasar molecular gas absorbers to determine this conversion factor $\alpha_{\rm [CI]} = M_{\rm mol} / L^\prime_{\rm [CI](1-0)}$ from the column density of H$_2$, which gives us the mass per unit column, and the [\ci]($J=1$) column density, which provides the luminosity per unit column. This technique allows us to make direct measurements of the relative abundances in high-redshift absorption-selected galaxies. Our sample spans redshifts of $z = 1.9 - 3.4$ and covers two orders of magnitude in gas-phase metallicity. We find that the [\ci]-to-$M_{\rm mol}$ conversion factor is metallicity dependent, with $\alpha_{\rm [CI]}$ scaling linearly with the metallicity: $\log \alpha_{\rm [CI]} = -1.13\times \log(Z/Z_{\odot}) + 1.33$, with a scatter of $\sigma_{\alpha_{\rm [CI]}} = 0.2$\,dex. Using a sample of emission-selected galaxies at $z\sim 0-5$, with both [\ci] and CO line detections, we apply the $\alpha_{\rm [CI]}$ conversion to derive independent estimates of the molecular gas mass and the CO-to-$M_{\rm mol}$, $\alpha_{\rm CO}$, conversion factor. We find a remarkable agreement between the molecular gas masses inferred from the absorption-derived $\alpha_{\rm [CI]}$ compared to typical $\alpha_{\rm CO}$-based estimates, which we confirm here to be metallicity-dependent as well, with an inferred slope that is consistent with $\alpha_{\rm [CI]}$ and previous estimates from the literature. These results thus support the use of the absorption-derived $\alpha_{\rm [CI]}$ conversion factor for emission-selected star-forming galaxies and demonstrate that both methods probe the same universal properties of molecular gas in the local and high-redshift universe. 
\end{abstract}


\section{Introduction} \label{sec:intro}

\begin{figure*}[!ht]
\centering
    \includegraphics[width=14cm]{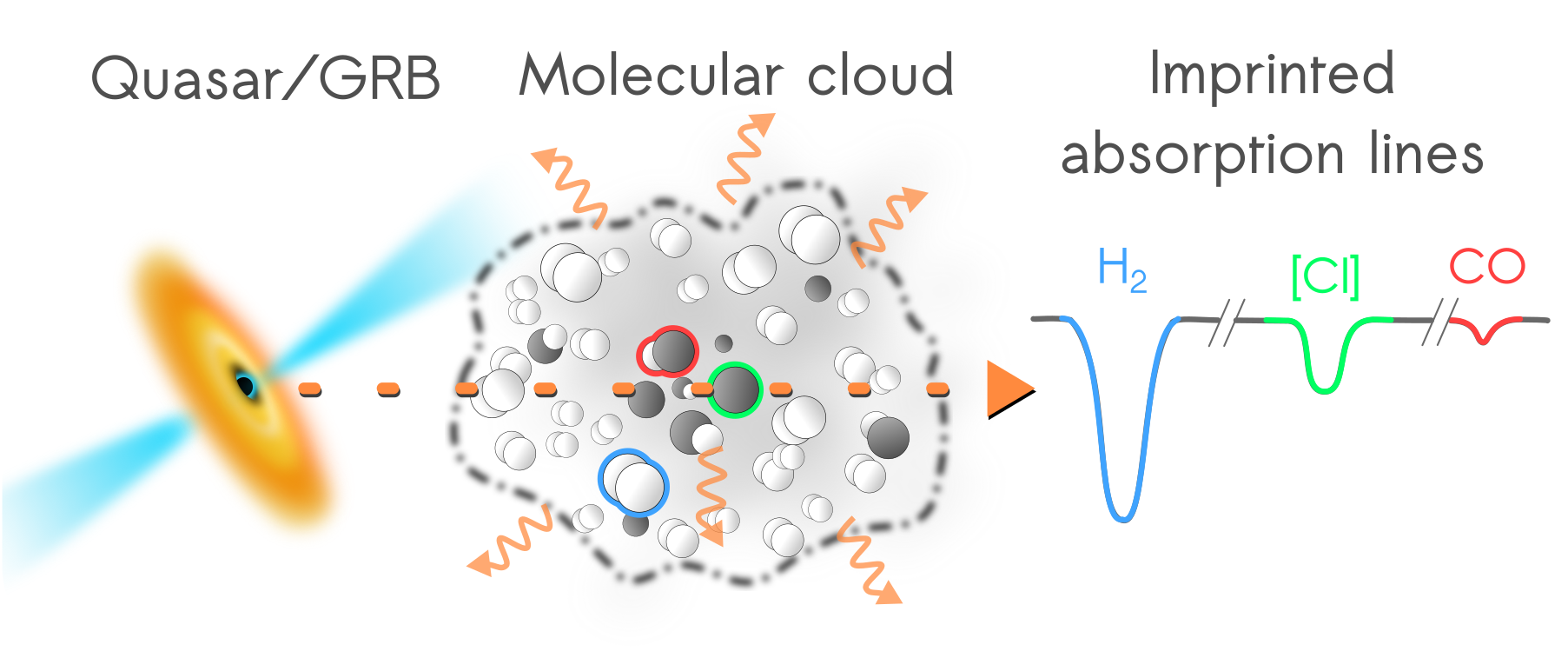}
    \caption{Schematic illustration of the use of GRBs and quasars as probes of molecular clouds in intervening or host galaxy absorption systems. The three most abundant molecular species and gas tracers are shown in blue (H$_2$), green ([\ci]), and red (CO). The excited fine-structure transitions of [\ci] and the rotational states of CO will emit light from the core of the molecular cloud, which can readily be detected in emission. As H$_2$ does not have a permanent dipole moment and the lowest rotational transition above the ground-state is only excited at temperatures far above that typically observed for the cold, molecular gas-phase, this molecule will only emit weakly, even though it is by far the most abundant. Measuring the relative column densities of these molecular species and gas tracers in absorption therefore provides an ideal probe of the relative abundances in molecular clouds.
    }
	\label{fig:illu}
\end{figure*}

Cold, molecular gas is the main fuel for star formation and is a vital component to studying the evolution of galaxies. However, the most abundant molecule, molecular hydrogen (H$_2$), cannot be detected routinely so other molecular gas tracers such as carbon monoxide (CO) and neutral atomic carbon ([\ci]) are used instead \citep[see][for a review]{Bolatto13}. The key to understanding star formation at high redshift is thus to advance these probes of the molecular content used locally to the high-redshift universe \citep{Solomon05,Carilli13}. 

The conversion between the CO or [\ci] luminosity to the total molecular gas mass have, however, only been directly constrained in Galactic molecular clouds and for a few local galaxies \citep{Solomon87,Bolatto13}. The main difficulty in expanding this relation to high redshift is that the molecular gas mass cannot, in the large majority of cases, be determined by estimating (for example) the cloud virial masses. Other approaches have therefore been applied, but common to them all is that they are based on a set of assumptions that have only been directly measured in the Milky Way or nearby galaxies, such as universal dust-to-gas ratio scaling relations \citep{Magdis11} or a similar high-redshift relation between gas surface-density and star formation \citep{Genzel12}. These are likely valid assumptions and the overall agreement between the absolute value and the metallicity evolution of these conversion factors inferred at high-redshift seems to validate the use of the locally derived scaling relations \citep{Daddi10,Genzel12,Magdis12}. However, this has not yet been directly verified.

In this Letter, we present a novel technique of estimating the conversion factor for the [\ci] luminosity to total molecular gas mass, $\alpha_{\rm [CI]}$, in high-redshift galaxies. While CO is still the most extensively surveyed species in emission \citep[e.g.][]{Tacconi13} and typically more abundant than [\ci] \citep{Ikeda02}, [\ci] has some advantages as a molecular gas tracer, especially at high redshifts \citep{Papadopoulos04a,Walter11,Valentino18}. 

For the approach presented here, we do not rely on any assumptions about the molecular content or locally derived scaling relations, but derive the conversion factor directly from observable quantities in gamma-ray burst (GRB) and quasar absorption-selected galaxies. We apply the absorption-derived $\alpha_{\rm [CI]}$ conversion factor to a recent sample of [\ci]-emission-detected galaxies and demonstrate the feasibility of this technique by comparing the inferred molecular gas masses to previous estimates relying on $\alpha_{\rm CO}$.

\setlength{\tabcolsep}{10pt}
\begin{deluxetable*}{lcccccr}\label{tab:sample}
\tablenum{1}
\tablecaption{Sample Properties of the GRB and Quasar Absorption-line Systems. }
\tablewidth{0pt}
\tablehead{
\colhead{Source} & \colhead{$z_{\mathrm{abs}}$} & \colhead{$\log N$(H$_2$)} & \colhead{$\log N$(\ci*)} & \colhead{$\log (Z/Z_{\odot})$} & \colhead{$\log \alpha_{\rm [CI]}$} & \colhead{Refs.}
}
\startdata
GRB DLAs \\
\noalign {\smallskip} \hline \noalign{\smallskip}
120815A & 2.358 & $20.42\pm 0.08$ & $13.86\pm 0.15$ & $-1.45\pm 0.03$ & $2.91\pm 0.16$ & (1,2) \\
121024A & 2.302 & $19.90\pm 0.17$ & $13.81\pm 0.10$ & $-0.76\pm 0.06$ & $2.20\pm 0.20$ & (1,3) \\
150403A & 2.057 & $19.90\pm 0.14$ & $14.72\pm 0.54$ & $-1.04\pm 0.04$ & $1.86\pm 0.21$ & (1,4) \\
181020A & 2.938 & $20.40\pm 0.04$ & $13.48\pm 0.06$ & $-1.57\pm 0.06$ & $3.06\pm 0.06$ & (4) \\
190114A & 3.376 & $19.45\pm 0.05$ & $13.27\pm 0.10$ & $-1.23\pm 0.07$ & $2.43\pm 0.16$ & (4) \\
\noalign {\smallskip} \hline \noalign{\smallskip}
QSO DLAs \\
\noalign {\smallskip} \hline \noalign{\smallskip}
J\,1513$+$0352 & 2.464 & $21.31\pm 0.01$ & $14.60\pm 0.06$ & $-0.84\pm 0.23$ & $2.82\pm 0.06$ & (5) \\
J\,0843$+$0221 & 2.786 &  $21.21\pm 0.02$ & $13.61\pm 0.02$ & $-1.52\pm 0.10$ & $3.71\pm 0.03$ & (6) \\
J\,0000$+$0048 & 2.526 & $20.43\pm 0.02$ & $15.54\pm 0.14$ & $0.46\pm 0.45$ & $1.00\pm 0.14$ & (7) \\
J\,2225$+$0527 & 2.133 & $19.40\pm 0.10$ & $13.80\pm 0.04$ & $-0.09\pm 0.05$ & $1.71\pm 0.11$ & (8) \\
J\,2140$-$0321 & 2.340 & $20.13\pm 0.07$ & $13.20\pm 0.04$ & $-1.05\pm 0.13$ & $3.04\pm 0.08$ & (9) \\
J\,0643$-$5041 & 2.659 & $18.54\pm 0.01$ & $12.47\pm 0.06$ & $-0.91\pm 0.09$ & $0.20\pm 0.23$ & (10) \\
J\,0816$+$1445 & 3.287 & $18.66\pm 0.27$ & $13.24\pm 0.02$ & $-1.10\pm 0.10$ & $1.53\pm 0.27$ & (11) \\
J\,1237$+$0647 & 2.690 & $19.21\pm 0.13$ & $14.54\pm 0.02$ & $0.34\pm 0.12$ & $0.78\pm 0.13$ & (12) \\
J\,1439$+$1117 & 2.418 & $19.38\pm 0.10$ & $14.02\pm 0.02$ & $0.16\pm 0.11$ & $1.47\pm 0.10$ & (13) \\
J\,0013$-$0029 & 1.973 & $18.86\pm 1.14$ & $13.37\pm 0.01$ & $-0.59\pm 0.05$ & $1.60\pm 0.33$ & (14,15) \\
J\,0528$-$2505 & 2.811 & $18.22\pm 0.12$ & $12.30\pm 0.10$ & $-0.91\pm 0.07$ & $2.03\pm 0.15$ & (14,15) \\
J\,0551$-$3638 & 1.962 & $17.42\pm 0.45$ & $13.33\pm 0.05$ & $-0.35\pm 0.08$ & $2.25\pm 0.12$ & (14,15) \\
J\,1232$+$0815 & 2.338 & $19.57\pm 0.10$ & $13.43\pm 0.07$ & $-1.43\pm 0.08$ & $2.50\pm 0.17$ & (14,15) \\
J\,1444$+$0126 & 2.087 & $18.16\pm 0.14$ & $12.77\pm 0.09$ & $-0.80\pm 0.09$ & $2.18\pm 0.06$ & (14,15) \\
\enddata
\tablecomments{{\bf References.} (1)~\citet{Bolmer19}; (2)~\citet{Kruhler13}; (3)~\citet{Friis15}; (4)~\citet{Heintz19}; (5)~\citet{Ranjan18}; (6)~\citet{Balashev17}; (7)~\citet{Noterdaeme17}; (8)~\citet{Krogager16}; (9)~\citet{Noterdaeme15}; (10)~\citet{Albornoz14}; (11)~\citet{Guimaraes12}; (12)~\citet{Noterdaeme10}; (13)~\citet{Srianand08}; (14)~\citet{Srianand05}; (15)~\citet{Noterdaeme08}.}
\end{deluxetable*}

\section{High-redshift [\ci]-to-$M_{\rm mol}$ conversion}

In rare cases, H$_2$ and [\ci] are detected in absorption-selected galaxies toward quasars \citep{Srianand05,Jorgenson10} and in GRB-host absorption systems \citep{Bolmer19,Heintz19}. The GRB-selected host galaxy absorption systems arguably provide the most robust estimate of the global [\ci]-to-H$_2$ abundance because they probe the central regions of their host galaxies, which we know for certain to be star-forming systems. The extremely strong quasar absorption systems (ES-DLAs), with neutral hydrogen column densities $N$(H\,{\sc i}) $> 10^{21.7}$\,cm$^{-2}$, are also proposed to probe similar low-impact parameters as GRB-host absorbers \citep{Noterdaeme14}. Therefore, these systems are similarly highlighted throughout. 

The advantage of detecting H$_2$ and [\ci] in absorption-selected galaxies is that the total and relative abundances of the molecular species and gas tracers can be directly measured for the molecular clouds (shown visually in Figure~\ref{fig:illu}), in galaxies at $z\sim 2-4$ during the peak of cosmic star formation. In this work, we extract all known GRB and quasar absorption-line systems with abundance measurements of both [\ci]($J=1$) and H$_2$ from our own work and the literature and list them in Table~\ref{tab:sample}. For each system, the absorption-derived redshift and gas-phase metallicity has also been reported as $\log (Z/Z_{\odot})$ = $\log N(\mathrm{X})/N(\mathrm{H}) - \log N(\mathrm{X})_{\odot}/N(\mathrm{H})_{\odot}$ relative to solar abundances \citep{Asplund09}. If multiple velocity components are detected, indicating individual clouds in the host absorption system, we sum the column densities for that system. We then derive the equivalent [\ci] luminosity to molecular gas mass ratio $\alpha_{\rm [CI]} = M_{\rm mol} / L^\prime_{\rm [CI](1-0)}$, for all absorbers in the sample, as follows. 

The spontaneous emission from the first excited state of neutral atomic carbon ([\ci]($J=1$); hence denoted as \ci*) gives rise to the line transition [\ci]$_{^3P_1-^3P_0}$ ([\ci]($1-0$)) at $\nu_{10} = 492.161$\,GHz in the rest frame. In terms of energy, the line luminosity is then given by $L'_{[\rm CI](1-0)} = h\, \nu_{10}\,A_{10}\,\Sigma_{[\rm CI^*]}$, where $h$ is Planck's constant, $\nu_{10}$ is the line frequency, $A_{10}=7.93\times 10^{-8}$\,s$^{-1}$ is the Einstein coefficient for this transition, and $\Sigma_{[\rm CI^*]}$ is the total population in the first excited state. Photons from stimulated emission might also contribute to the emitted light, but because the molecular gas is completely shielded from the far-infrared (FIR) background radiation we argue that this component is negligible. We then calculate the total molecular gas mass, which we define as $M_{\rm mol} = M_{\rm H_2}\times f_{\rm m}$, where $f_{\rm m} = 1.36$ to include the contribution from helium and heavier elements confined in the molecular region. The metallicity has only a small effect ($\lesssim1$\%) on this number. Here, $M_{\rm H_2} = m_{\rm H_2}\Sigma_{\rm H_2}$, where $m_{\rm H_2}$ is the mass of a single hydrogen molecule and $\Sigma_{\rm H_2}$ is the total population of molecular hydrogen.

We can then define the total molecular gas mass to the equivalent [\ci]($1-0$) line luminosity ratio as
\begin{equation}\label{eq:lcimolmass}
\frac{M_{\rm mol}}{L'_{[\rm CI](1-0)}} = \frac{m_{\rm H_2}\,f_{\rm m}\,\Sigma_{\rm H_2}}{h\,\nu_{10}\,A_{10}\,\Sigma_{[\rm CI^*]}}~.
\end{equation}
For GRB and quasar absorbers, we directly measure the column density of elements in the line of sight. The column density ratio of $N$(H$_2$) to the first excited transition of neutral atomic carbon $N$(\ci*) must therefore be the same as the relative total number population, so $N$(H$_2$)/$N$(\ci*) = $\Sigma_{\rm H_2}/\Sigma_{[\rm CI^*]}$. Substituting this into Eq.~\ref{eq:lcimolmass} yields
\begin{equation}\label{eq:colmassh2}
   \frac{M_{\rm mol}}{L'_{[\rm CI](1-0)}} = \frac{{N(\rm H_2)}}{N({\rm CI^*})}\,\frac{m_{\rm H_2}\,f_{\rm m}}{h\,\nu_{10}\,A_{10}}~.
\end{equation}

Converting these column-density-measured abundance ratios to more typical units, this then becomes
\begin{equation}
    \frac{M_{\rm mol}}{L'_{[\rm CI](1-0)}} 
    = 1.30\times10^{-4}\frac{N({\rm H_2})}{N({\rm CI^*})}\,
    \mathrm{M}_{\odot}\,\mathrm{(K\,km\,s^{-1}\,pc^2)}^{-1},
\end{equation}

Following the same approach we could in principle also derive the equivalent second excited [\ci]($2-1$) line luminosity to molecular gas mass ratio. 
However, the column density of the second excited fine-structure transition is often not well-constrained in GRB and quasar absorption-line analyses, so we here only provide the results for $M_{\rm mol}/L'_{[\rm CI](1-0)}$.  \\

\section{Results}

\begin{figure}[!t]
    \includegraphics[width=\columnwidth]{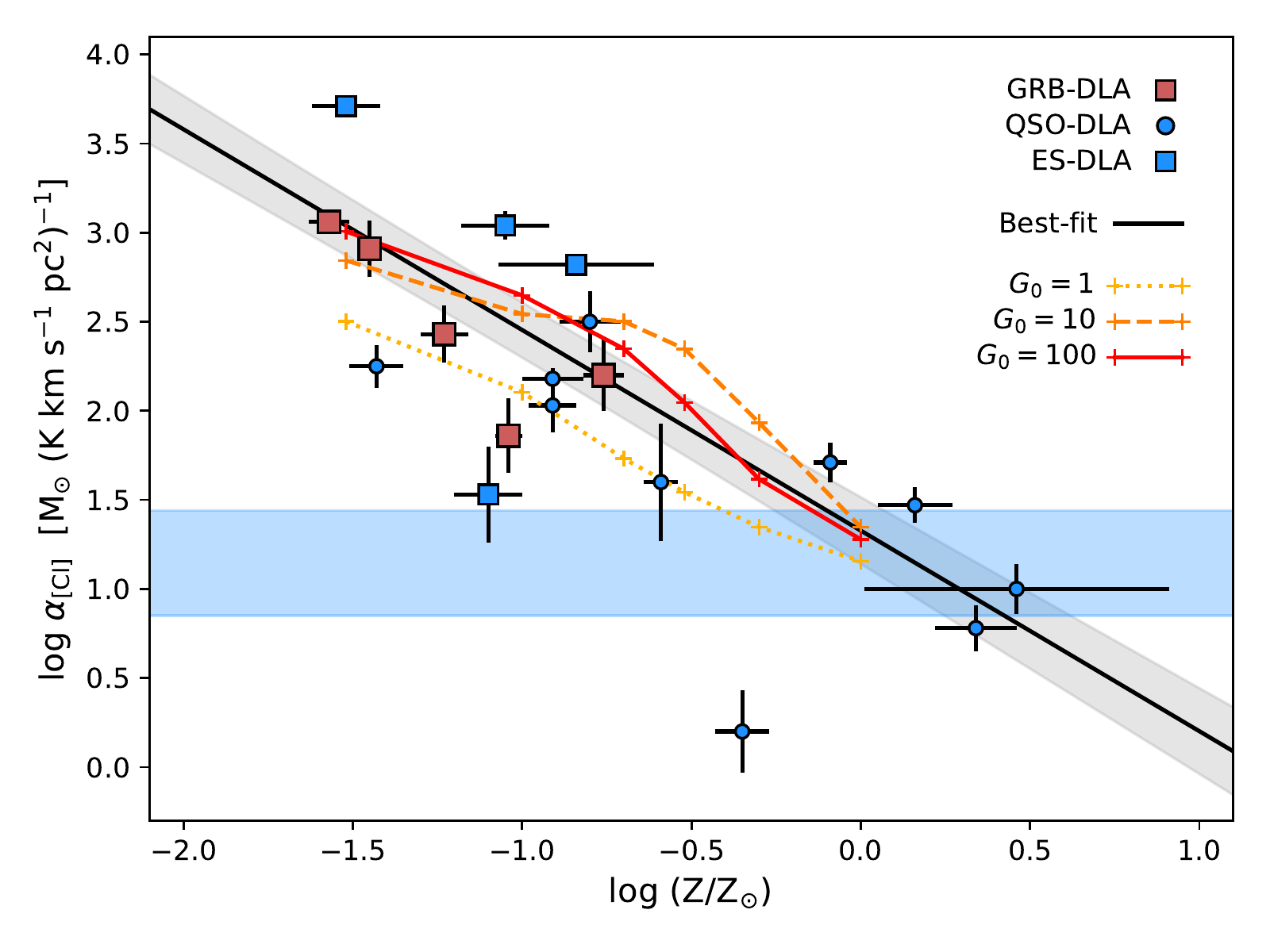}
    \caption{Absorption-derived metallicity evolution of the molecular gas mass to [\ci] line luminosity. The red squares denote GRB absorption systems, and the blue symbols represent absorbers in quasar (QSO) sightlines, where the large squares denote the extremely strong DLAs (ES-DLAs). The best-fit linear relation $\log \alpha_{\rm [CI]} = -1.13\times \log(Z/Z_{\odot}) + 1.33$ is shown by the black solid line and the error on the fit is shown by the gray shaded area. The metallicity evolution of $\alpha_{\rm [CI]}$ for a range of radiation field strengths determined from numerical hydrodynamical simulations by \cite{Glover16} is shown for comparison. The blue shaded region shows the range of $\alpha_{\rm [CI]}$ estimated for local galaxies \citep{Crocker19} or high-redshift galaxies \citep{Valentino18} using the average Galactic conversion factor of $\alpha_{\rm CO,MW}=4.3\,M_{\odot}$\,(K\,km\,s$^{-1}$\,pc$^2)^{-1}$ \citep{Bolatto13}. 
    }
	\label{fig:cih2hm}
\end{figure}

The measurements of $\alpha_{\rm [CI]}$ for the GRB and quasar absorption-selected galaxies are provided Table~\ref{tab:sample} and shown as a function of metallicity in Figure~\ref{fig:cih2hm}. We find that $\alpha_{\rm [CI]}$ increases with decreasing metallicity, with a best-fit relation of 
\begin{equation}
    \log \alpha_{\rm [CI]} = (-1.13\pm 0.19)\times \log(Z/Z_{\odot}) + (1.33\pm 0.21)
\end{equation}
The observed scatter is likely dominated by variations in the physical properties of each molecular cloud (such as temperature and density) and the intensity of the ultraviolet (UV) background field. For comparison, we show the metallicity evolution of $\alpha_{\rm [CI]}$ as inferred from recent numerical hydrodynamical simulations for a range of UV radiation field strengths by \cite{Glover16}. 
The linear metallicity relation of $\alpha_{\rm [CI]}$ found here matches well with these theoretical expectations. This result is also reflected in the total [\ci]-to-H$_2$ column density ratio, for which we find a best-fit, metallicity-dependent linear relation of 
\begin{equation}
    \log N({\rm CI/H}_2) = (1.06\pm 0.23)\times \log(Z/Z_{\odot}) - (4.79\pm 0.25)
\end{equation}
from the same sample of absorption-selected galaxies. This is consistent with the typically adopted constant abundance ratio of $X_{\rm [CI]} = 3\times 10^{-5}$ \citep{Papadopoulos04b}, but only at solar metallicities.

As an example, from this relation we estimate that galaxies with stellar masses of $\log (M_{\star}/M_{\odot}) \approx 10.5$ at $z\sim 2.2$ \citep[corresponding to solar metallicity abundances following the observed mass-metallicity relation;][]{Maiolino08} would have a conversion factor of $\log (M_{\rm mol}/M_{\odot}) = \log L'_{\rm [CI](1-0)} + (1.33\pm 0.21)$ using our calibration. This is consistent with the observed ratios between $L'_{\rm [CI](1-0)}$ and molecular gas masses determined for local galaxies \citep{Crocker19} and high-redshift galaxies \citep{Valentino18} using the Milky Way conversion factor of $\alpha_{\rm CO} = 4.3\,M_\odot$\,(K\,km\,s$^{-1}$\,pc$^2$)$^{-1}$ \citep{Bolatto13}, shown as the blue shaded region in Figure~\ref{fig:cih2hm}. We are thus able to reproduce the average Milky Way $\alpha_{\rm [CI]}$ and $X_{\rm [CI]}$ conversion factors at solar metallicities with this approach. \\

\begin{figure}[!h]
	\includegraphics[width=9cm]{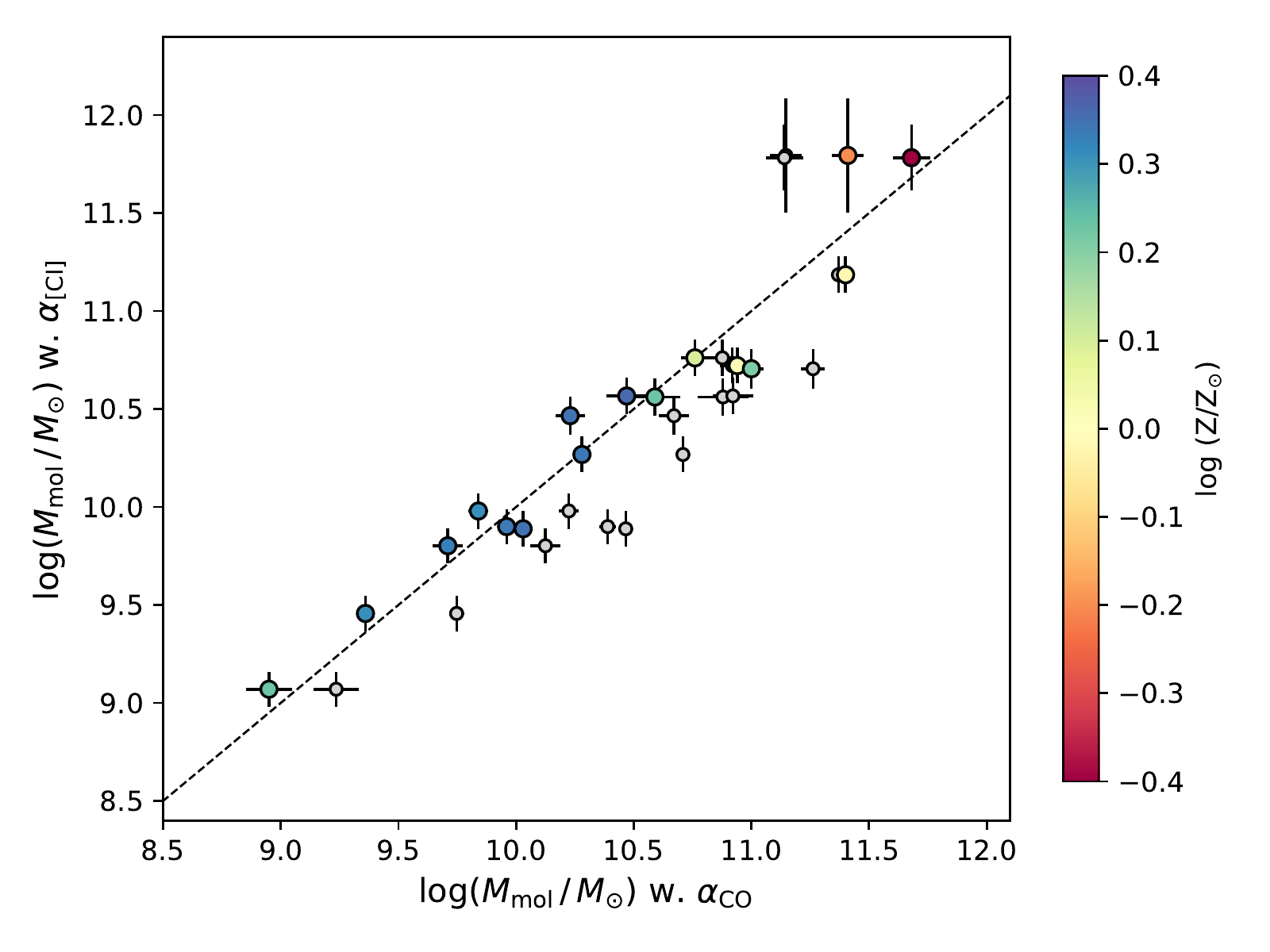}
	\caption{Comparison between $\alpha_{\rm [CI]}$- and $\alpha_{\rm CO}$-derived molecular gas masses. The symbols show all the [\ci]-emitting galaxies selected from the compilation by \cite{Valentino18}. The colored points show the estimated molecular gas mass assuming a metallicity-dependent conversion factor of $\alpha_{\rm CO} = \alpha_{\rm CO,MW} \times 10^{-1.27\log(Z/Z_{\odot})}$ from \cite{Genzel12}, where the gray points represent the same galaxies but assuming a constant conversion factor of $\alpha_{\rm CO,MW} = 4.3\,M_{\odot}$\,(K\,km\,s$^{-1}$\,pc$^2)^{-1}$ (therefore only offset along the $x$-axis). The points are color-coded as a function of gas-phase metallicity, determined from either the fundamental plane \citep{Mannucci10} or a redshift-dependent mass-metallicity relation \citep{Maiolino08}. The dashed line shows the line of equality. }
	\label{fig:mmolcomp}
\end{figure}

\section{Comparison to $\alpha_{\rm CO}$}

To verify the results and demonstrate the application of our new absorption-derived calibration of $\alpha_{\rm [CI]}$ further, we use the recent sample of emission-selected galaxies with positive detections of [\ci] compiled by \cite{Valentino18}. This sample includes local, intermediate- and high-redshift galaxies and quasars spanning a large redshift range of $z\sim 0-5$. We removed the sources with significant active galactic nucleus (AGN) contribution to only consider regular star-forming galaxies in our analysis. To infer the gas-phase metallicities for these galaxies, we use the fundamental relation \citep[FMR;][]{Mannucci10} if both the stellar mass and star-formation rate are available (which is only the case for the galaxies at $z\sim 1.2$), or a redshift-dependent mass-metallicity relation \citep{Maiolino08} for the rest. While there is some uncertainty related to these scaling relations we note that varying the stellar mass at a given redshift by $\sim 0.5$\,dex only causes a shift in the inferred metallicity of $\sim 0.1$\,dex. For the large majority of the galaxies at $z\gtrsim 2.5$ in this sample, only dust masses have been derived. So, for these we inferred the stellar mass by assuming a redshift-dependent scaling with the dust mass \citep{Calura17}. These are not included in the main analysis, but are highlighted where appropriate.

We then apply our $\alpha_{\rm [CI]}$ relation to calculate the molecular gas mass for these galaxies. In Figure~\ref{fig:mmolcomp} we compare our measurements to those derived using $L^{\prime}_{\rm CO(1-0)}$, assuming either the metallicity-dependent conversion factor from \cite{Genzel12} or a constant Galactic conversion factor of $\alpha_{\rm CO, MW} = 4.3\,M_{\odot}$\,(K\,km\,s$^{-1}$\,pc$^2)^{-1}$ \citep{Bolatto13}. The inferred molecular gas masses, using our absorption-derived calibration of $\alpha_{\rm [CI]}$ or the metallicity-dependent $\alpha_{\rm CO}$ conversion, follow each other closely through three orders of magnitude in molecular gas mass. 
The comparison with $\alpha_{\rm CO, MW}$ indicates that assuming an average Galactic [\ci]- or CO-to-H$_2$ constant scaling may systematically over-predict the molecular gas mass of massive galaxies at $z\sim2$, while under-predicting the amount of molecular gas in low-mass, low-metallicity galaxies that dominate the high-redshift galaxy population.

\begin{figure}[!t]
	\includegraphics[width=9cm]{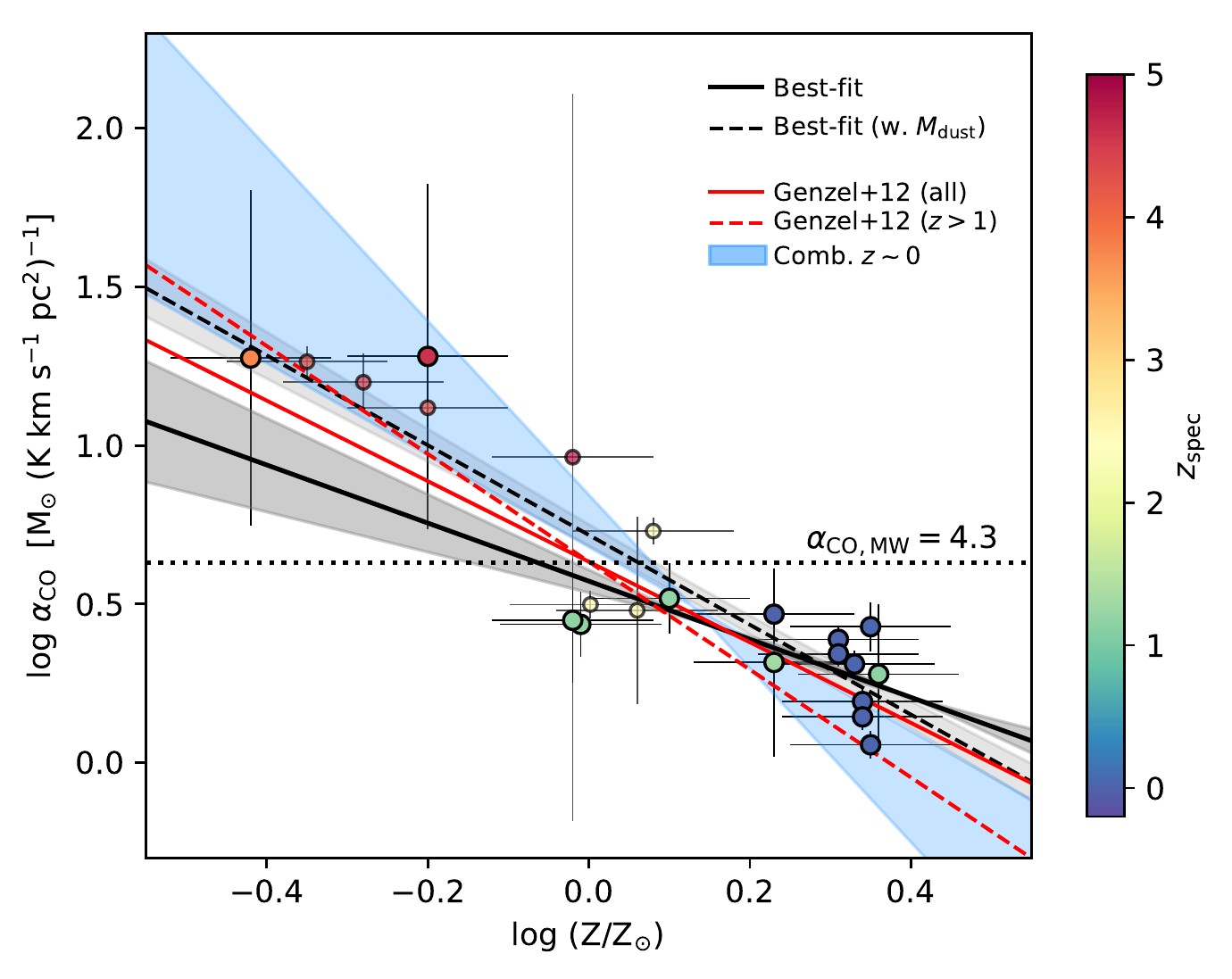}
	\caption{Metallicity evolution of $\alpha_{\rm CO}$ for a sample of $0 < z < 5$ CO- and [\ci]-emitting galaxies. The values of $\alpha_{\rm CO}$ shown here are derived from the CO($1-0$) and [\ci]($1-0$) line luminosities, the $\alpha_{\rm [CI]}$ conversion and the inferred gas-phase metallicities, as in Figure~\ref{fig:mmolcomp}. The points are color-coded by redshift. The solid black line shows the best-fit linear relation which has a slope $\alpha_{\rm CO} \propto (Z/Z_{\odot})^{-0.93\pm 0.28}$ and intercept at $Z=Z_\odot$ of $\alpha_{\rm CO} = (3.8^{+0.9}_{-0.7})\,M_{\odot}$\,(K\,km\,s$^{-1}$\,pc$^2)^{-1}$. In the plot we also include estimates of $\alpha_{\rm CO}$ for the subset of $z\gtrsim 2.5$ galaxies where the stellar mass was inferred assuming a redshift-dependent scaling with the dust mass. Including these points in the fit yield a slightly steeper linear relation. For comparison, we include independent estimates of the metallicity evolution of $\alpha_{\rm CO}$ at high-$z$ from \cite{Genzel12} and show a range of $z\sim 0$ estimates \citep[combining][]{Israel97,Leroy11,Amorin16}. }
	\label{fig:alphaco}
\end{figure}

We then apply our $\alpha_{\rm [CI]}$ calibration to also examine CO as a molecular gas mass tracer in an independent way. The metallicity dependence of $\alpha_{\rm CO}$ is expected to be similar to that of $\alpha_{\rm [CI]}$ because [\ci] probes the same, more shielded inner regions of molecular clouds as CO \citep{Ikeda02}. In Figure~\ref{fig:alphaco} we again show the absorption-derived values of $M_{\rm mol}$ using our $\alpha_{\rm [CI]}$ calibration, but now relative to $L^{\prime}_{\rm CO(1-0)}$ measured for a large subset of the [\ci]-emitting galaxies, as a function of the inferred gas-phase metallicity. We find that the metallicity evolution of $\alpha_{\rm CO}$ is best fit with a slope of $\alpha_{\rm CO} \propto (Z/Z_{\odot})^{-0.93\pm 0.28}$ and an intercept at solar metallicities $Z=Z_\odot$ \citep[equal to $12 + \log$(O/H) $=8.69$;][]{Asplund09} of $\alpha_{\rm CO} = (3.8^{+0.9}_{-0.7})\,M_{\odot}$\,(K\,km\,s$^{-1}$\,pc$^2)^{-1}$, consistent with the metallicity evolution of $\alpha_{\rm [CI]}$ and the constant and metallicity-dependent values of $\alpha_{\rm CO}$ estimated for the Milky Way and local galaxies \citep{Israel97,Bolatto13}. Including the more uncertain measurements based on the dust mass for the galaxies at $z\gtrsim 2.5$, we find a slightly steeper metallicity evolution with slope $\alpha_{\rm CO} \propto (Z/Z_{\odot})^{-1.41\pm 0.13}$ and intercept at $Z=Z_\odot$ of $\alpha_{\rm CO} = (5.2^{+1.2}_{-0.9})\,M_{\odot}$\,(K\,km\,s$^{-1}$\,pc$^2)^{-1}$, though more consistent with previous estimates.
We caution that for these estimates of $\alpha_{\rm CO}$ we rely on a set of assumptions that were not required to derive $\alpha_{\rm [CI]}$ and the results are therefore only meant as indicative of the actual underlying scaling relations. 

\section{Discussion}

The remarkable consistency between local estimates and the method presented here for high-redshift galaxies, based only on the mass-derived metallicity and the $L^\prime_{\rm [CI](1-0)}$ measurements, provides further validation of the absorption-derived calibration of $\alpha_{\rm [CI]}$. These consistent scaling relations also suggest that the molecular gas properties and cloud compositions are uniform at all redshifts, on average, for a given metallicity. Moreover, they support the scenario where both the absorption and emission measured properties can be used as representative probes of the molecular gas-phase. Finally, it also lends credibility to the use of simple galaxy mass-metallicity scaling relations to infer more accurate molecular gas mass conversion factors for galaxies at low and high redshifts. Combining our measurement of $\alpha_{\rm CO}$ with previous literature measurements suggests a common metallicity-dependent conversion factor of $\alpha_{\rm CO} = 4.5 \times (Z/Z_{\odot})^{-1.4}$ for star-forming galaxies at all redshifts. 

While GRB and quasar absorption-selected galaxies by selection do not probe the same field galaxies as emission-selected surveys, there is recent evidence that they sample the same underlying star-forming galaxy population, though typically at the faint, low-mass end \citep{Kruhler15,Krogager17}. Independent of the nature of the galaxies, simulations seem to suggests that the metallicity is the primary factor regulating the abundances of [\ci] or CO relative to H$_2$ in molecular clouds \citep{Wolfire10,Bolatto13,Glover16}. We measure and account for the metallicity of individual systems directly in the absorption-derived \(\alpha_{\rm [CI]}\) relation. Absorbing galaxies typically also only show small variations in gas-phase metallicity internally throughout their interstellar medium \citep[ISM;][]{Izzo17}, of order 0.02\,dex\,kpc$^{-1}$ \citep{Christensen14}. These considerations and the overall excellent agreement between the physical properties derived for molecular clouds in absorption and emission suggests that the line-of-sight measurements thus provide a good estimate of the average relative abundances of the molecular gas tracers in the probed environments of high-redshift star-forming galaxies.

\section{Summary and outlook}

In conclusion, while the local, constant conversion factors are an adequate approximation within a factor of about two for the higher mass, and hence approximately solar metallicity galaxies detected in current high-redshift galaxy surveys, these should not be applied uniformly. More accurate theoretical simulations of galaxies and further samples of fainter galaxies detected with mm/sub-mm observatories and soon with the next generation of telescopes, such as the {\it James Webb Space Telescope}, now require more accurate estimates of the molecular gas mass content for low-metallicity, low-mass galaxies at all redshifts. This work provides the first direct calibration of the conversion between the [\ci] line luminosity $L'_{\rm [CI](1-0)}$ to the total molecular gas mass for high-redshift galaxies. This calibration is also specifically sensitive to the metal-poor regime that is difficult to survey at low redshifts, but is now increasingly being probed by sub-mm observations of low-mass galaxies that dominate the high-redshift galaxy population \citep{Hashimoto19}. The remarkable consistency between the scaling relations derived here for the absorption-selected galaxies, compared to that observed for molecular clouds in the Milky Way and local galaxies, validates this technique and suggests that the properties of molecular clouds are ubiquitous both in the local and high-redshift universe. This universality means that the metallicity-dependent $\alpha_{\rm [CI]}$ and $\alpha_{\rm CO}$ conversion factors can be applied at all redshifts.

\acknowledgments

First, we would like to thank the referee for a clear and constructive report, greatly improving the presentation of the results in this Letter. We would also like to thank Francesco Valentino, Georgios E. Magdis, Johan P. U. Fynbo and P\'{a}ll Jakobsson for enlightening discussions during the early stages of this work. Finally, we would like to thank H\'eloïse F. Stevance (\url{https://hfstevance.com}) for providing us with Figure 1 that efficiently illustrates the approach presented in this Letter. K.E.H. acknowledges support by a Project Grant (162948--051) from The Icelandic Research Fund. D.W. is supported by Independent Research Fund Denmark grant DFF -- 7014-00017. The Cosmic Dawn Center is supported by the Danish National Research Foundation under grant No.\ 140.

\software{astropy \citep{Astropy},  
          emcee \citep{Emcee},
          numpy \citep{Numpy}, 
          lmfit \citep{LMfit},
          matplotlib \citep{Matplotlib}
          }

\bibliography{ref}
\bibliographystyle{aasjournal}

\end{document}